\documentclass[aps,prl,reprint,superscriptaddress]{revtex4-1}

\usepackage{graphicx}
\usepackage{dcolumn}
\usepackage{natbib}
\usepackage{color}
\usepackage{amssymb,amsmath}
\usepackage{hyperref}
\usepackage{lipsum}

\begin{document}
	
	\title{A Unified Hamiltonian Formulation for Energy Loss, Entropy Evolution, and Fusion Performance in Plasmas}
	
	\author{J. S. Finberg}
	\email{jsf2178@columbia.edu}
	\affiliation{Columbia University Department of Physics, New York, NY, USA}
	
	\date{\today}
	
	\begin{abstract}
		We develop a comprehensive Hamiltonian formulation for plasma dynamics that unifies collisionless (gyrokinetic) and collisional processes. Our framework rigorously describes the evolution of free energy and entropy during the transition from Maxwellian to non-Maxwellian distributions, explicitly coupling microscopic turbulent processes with macroscopic measures of energy confinement and fusion performance. Unlike standard gyrokinetic treatments that treat collisions as a minor perturbation, our approach incorporates a collision operator directly into the Hamiltonian structure, thereby accounting for irreversible dissipation and entropy production. We derive quantitative relations linking turbulence intensity, entropy production, energy confinement time, and fusion yield. Our work builds on recent energetic bounds and optimal mode analyses by Helander and Plunk \cite{HelanderPlunk2022a} \cite{PlunkHelander2022b} \cite{PlunkHelander2023} and on Zhdankin’s generalized entropy production framework \cite{Zhdankin2022}, thus providing a bridge between microscopic kinetics and reactor-scale performance.
	\end{abstract}
	
	\maketitle
	
	%%%%%%%%%%%%%%%%%%%%%%%%%%%%%%%%%%%%%%%%%%%%%%%%%%%%%%%%%%%%%%%%%%%%%%
	%% I. Introduction
	%%%%%%%%%%%%%%%%%%%%%%%%%%%%%%%%%%%%%%%%%%%%%%%%%%%%%%%%%%%%%%%%%%%%%%
	
	\section{Introduction}
	
	Fusion energy research requires a deep understanding of plasma confinement, energy transport, and entropy evolution. In a well-confined plasma, the equilibrium distribution is typically Maxwellian, representing a state of maximum entropy under the constraints of energy and particle conservation. However, various plasma processes such as external heating, steep gradients, or fast particle injection drive the system away from this equilibrium, leading to non-Maxwellian features that store free energy. This free energy can then excite microinstabilities and turbulence, which degrade confinement by enhancing transport.
	
	Gyrokinetic theory has long been a cornerstone for describing turbulence in magnetized plasmas. By averaging over the fast cyclotron motion, the full Vlasov--Maxwell system reduces to a set of gyrokinetic equations that capture the slow dynamics responsible for turbulent transport \cite{FriemanChen1982}. Over recent years, Helander and Plunk have provided rigorous energetic bounds on the growth of free energy in these gyrokinetic systems \cite{HelanderPlunk2022a}. In subsequent studies, Plunk and Helander extended this work to develop an optimal mode analysis that identifies the perturbations which maximize free energy growth \cite{PlunkHelander2022b} \cite{PlunkHelander2023}. Such analyses constrain the maximum turbulence intensity achievable in fusion plasmas and help explain why, despite large drive, the system exhibits a bounded response.
	
	Complementary to these studies, Zhdankin \cite{Zhdankin2022} constructed a generalized framework for quantifying entropy production in collisionless plasmas. By employing Casimir invariants of the Vlasov equation, Zhdankin demonstrated that even in the absence of explicit collisionality, phase-space mixing (or turbulent cascades) can lead to effective entropy production. This result implies that irreversible processes occur in plasmas through turbulent mixing, even if the underlying Hamiltonian dynamics conserve the fine-grained Boltzmann entropy.
	
	While these works have significantly advanced our understanding of microinstabilities and turbulence, a gap remains in directly linking these microscopic phenomena to macroscopic reactor performance metrics such as the energy confinement time ($\tau_E$) and fusion reaction rate ($R_f$). Moreover, many traditional approaches treat collisions as a minor, secondary effect. In contrast, our work integrates collisions into the Hamiltonian framework to fully capture irreversible energy loss and entropy evolution. This unified approach allows us to derive a quantitative relationship between the turbulent drive (characterized by a growth rate $\lambda$), the effective entropy production (denoted by $S$), and the energy confinement time, with the result that
	\[
	\tau_E \sim \frac{1}{\tau\,\lambda\,S}\,,
	\]
	where $\tau$ is a characteristic nonlinear timescale.
	
Our formulation corrects earlier misconceptions that improved confinement (i.e., longer $\tau_E$) would reduce fusion yield; rather, a longer $\tau_E$ is directly linked to higher fusion performance. This insight provides a clear target for experimental efforts aimed at turbulence suppression, for instance via $E \times B$ shear stabilization \cite{CandyWaltz2003} and optimized equilibrium profiles \cite{Barnes2011}. The primary contributions of this work are as follows. We develop a unified Hamiltonian formulation that combines reversible (collisionless) gyrokinetic dynamics with irreversible collisional effects. We present a rigorous derivation showing how turbulence-driven free energy cascades lead to effective entropy production, building on the frameworks established by Helander and Plunk \cite{HelanderPlunk2022a} \cite{PlunkHelander2022b} \cite{PlunkHelander2023} and Zhdankin \cite{Zhdankin2022}. We also establish quantitative relations linking turbulence intensity ($\lambda$), entropy production ($S$), energy confinement time ($\tau_E$), and fusion performance ($R_f$). Finally, we discuss the practical implications of minimizing turbulent entropy production for achieving improved confinement in fusion reactors. In the following sections, we develop these ideas in detail, starting with a full presentation of the Hamiltonian description and entropy evolution (§II), then discussing gyrokinetic turbulence (§III), and finally deriving the impact on energy confinement and fusion performance (§IV). We conclude in §V with a summary and outlook.

%%%%%%%%%%%%%%%%%%%%%%%%%%%%%%%%%%%%%%%%%%%%%%%%%%%%%%%%%%%%%%%%%%%%%%
%% II. Hamiltonian Formulation and Entropy Evolution (Part 2)
%%%%%%%%%%%%%%%%%%%%%%%%%%%%%%%%%%%%%%%%%%%%%%%%%%%%%%%%%%%%%%%%%%%%%%

\section{Hamiltonian Formulation and Entropy Evolution}

In this section we present a detailed derivation of the Hamiltonian formulation for plasma dynamics, derive the collisionless kinetic equation, and discuss the evolution of entropy. We also describe how irreversible effects are introduced via a collision operator.

\subsection{Hamiltonian Description of Plasma Dynamics}

We consider a plasma consisting of $N$ charged particles with positions $\mathbf{x}_i$, momenta $\mathbf{p}_i$, masses $m_i$, and charges $q_i$, interacting via self-consistent electromagnetic fields. The total energy of the system is given by the Hamiltonian
\begin{equation}
	H = \sum_{i=1}^{N} \left[ \frac{p_i^2}{2m_i} + q_i\,\phi(\mathbf{x}_i) + q_i\,\mathbf{v}_i\cdot\mathbf{A}(\mathbf{x}_i) \right] + \frac{1}{2}\int \left( \frac{\mathbf{B}^2}{\mu_0} + \epsilon_0\,\mathbf{E}^2 \right) d^3x \,,
	\label{eq:Hamiltonian}
\end{equation}
where $\phi(\mathbf{x})$ and $\mathbf{A}(\mathbf{x})$ are the scalar and vector potentials, respectively; the electric and magnetic fields are defined by $\mathbf{E} = -\nabla\phi - \partial_t \mathbf{A}$ and $\mathbf{B} = \nabla\times\mathbf{A}$.

Hamilton's equations, together with Liouville's theorem, guarantee that the phase-space volume is preserved during the time evolution of the system \cite{Morrison1998}. This preservation is central to the derivation of the kinetic (or Vlasov) equation. In fact, the evolution of the single-particle distribution function $f(\mathbf{x},\mathbf{v},t)$ is governed by
\begin{equation}
	\frac{\partial f}{\partial t} + \mathbf{v}\cdot\nabla_{\mathbf{x}} f + \frac{q}{m}\left( \mathbf{E} + \mathbf{v}\times\mathbf{B} \right)\cdot\nabla_{\mathbf{v}} f = 0 \,.
	\label{eq:Vlasov}
\end{equation}
Equation \eqref{eq:Vlasov} is derived by following the characteristics defined by Hamilton's equations, and it expresses the conservation of $f$ along particle trajectories. For a comprehensive review of the Hamiltonian formulation in plasma dynamics, see \cite{Morrison1998}.

In the presence of a strong magnetic field, one often performs a gyroaveraging procedure to obtain the gyrokinetic equations. This involves averaging over the fast cyclotron motion, leading to a reduced description that focuses on the slower drift dynamics \cite{FriemanChen1982}. Although our starting point is Eq.~\eqref{eq:Vlasov}, the gyrokinetic model retains the essential Hamiltonian structure while reducing the dimensionality of the problem.

\subsection{Entropy in Collisionless and Collisional Plasmas}

The Boltzmann entropy for the plasma is defined as
\begin{equation}
	S[f] = - k_B \int f(\mathbf{x},\mathbf{v},t)\,\ln\!\big(f(\mathbf{x},\mathbf{v},t)\big)\,d^3x\,d^3v \,.
	\label{eq:Entropy}
\end{equation}
In an ideal, collisionless plasma governed by Eq.~\eqref{eq:Vlasov}, the fine-grained entropy $S[f]$ is conserved. This is a direct consequence of Liouville's theorem: as the distribution function is advected in phase space without distortion of the phase-space volume, the entropy remains constant.

However, in realistic plasmas, the situation is more complex. Even in systems where collisions are very weak, turbulent phase-space mixing (or filamentation) can occur. As argued by Zhdankin \cite{Zhdankin2022}, although the fine-grained entropy remains constant, a coarse-grained view of the distribution function shows an effective increase in entropy. This is because turbulent mixing transfers free energy from macroscopic scales to ever finer scales, which are eventually smoothed out by even weak collisions. Thus, an increase in entropy is observed in practice, and this effective entropy production is a signature of irreversible processes.

\subsection{Inclusion of Collisions: The Role of the Collision Operator}

To capture irreversible processes and the associated entropy production, we extend the collisionless Vlasov equation by incorporating a collision operator, $C(f)$. The kinetic equation then becomes
\begin{equation}
	\frac{\partial f}{\partial t} + \mathbf{v}\cdot\nabla_{\mathbf{x}} f + \frac{q}{m}\left( \mathbf{E}+\mathbf{v}\times\mathbf{B} \right)\cdot\nabla_{\mathbf{v}} f = C(f) \,.
	\label{eq:Boltzmann}
\end{equation}
For a fully ionized plasma, the Landau collision operator is often employed. This operator, originally formulated by Landau \cite{Landau1936_terHaar}, describes the small-angle Coulomb collisions between charged particles. It has the crucial property of conserving the number of particles, momentum, and energy, while ensuring that the entropy production satisfies
\begin{equation}
	\frac{dS}{dt} \ge 0 \,.
\end{equation}
Thus, collisions drive the distribution function towards a Maxwellian, increasing the entropy as free energy is dissipated. In the Hamiltonian picture, the reversible dynamics (given by Eqs.~\eqref{eq:Hamiltonian} and \eqref{eq:Vlasov}) are supplemented by the irreversible processes introduced via $C(f)$ in Eq.~\eqref{eq:Boltzmann}.

In our unified formulation, both the collisionless evolution (which preserves phase-space volume) and the collisional dissipation (which produces entropy) are accounted for. This dual treatment allows us to examine how turbulence-induced deviations from Maxwellian equilibrium are ultimately resolved by collisional processes, leading to irreversible energy loss and a rise in entropy.

%%%%%%%%%%%%%%%%%%%%%%%%%%%%%%%%%%%%%%%%%%%%%%%%%%%%%%%%%%%%%%%%%%%%%%
%% End of Part 2
%%%%%%%%%%%%%%%%%%%%%%%%%%%%%%%%%%%%%%%%%%%%%%%%%%%%%%%%%%%%%%%%%%%%%%
%%%%%%%%%%%%%%%%%%%%%%%%%%%%%%%%%%%%%%%%%%%%%%%%%%%%%%%%%%%%%%%%%%%%%%
%% III. Gyrokinetic Turbulence and Free Energy Transport (Part 3)
%%%%%%%%%%%%%%%%%%%%%%%%%%%%%%%%%%%%%%%%%%%%%%%%%%%%%%%%%%%%%%%%%%%%%%

\section{Gyrokinetic Turbulence and Free Energy Transport}

\subsection{Gyrokinetic Approximation and Turbulence}
In a strongly magnetized plasma, the fast gyromotion of charged particles allows for a simplification of the kinetic description. By performing an averaging procedure over the fast cyclotron motion, one obtains the gyrokinetic equations that describe the evolution of the gyrocenter distribution function $g(\mathbf{R}, v_\parallel, \mu, t)$ in a reduced (five-dimensional) phase space \cite{FriemanChen1982}. This reduction retains the essential physics of low-frequency dynamics, including the $\mathbf{E}\times \mathbf{B}$ drift, magnetic trapping, and wave--particle interactions, while eliminating the fast timescales associated with cyclotron motion.

The gyrokinetic equation can be schematically written as
\begin{equation}
	\frac{\partial g}{\partial t} + v_\parallel \frac{\partial g}{\partial l} + \mathbf{v}_d \cdot \nabla g + \mathcal{N}(g) = C(g) \,,
	\label{eq:gyrokinetic}
\end{equation}
where $v_\parallel$ is the velocity parallel to the magnetic field, $l$ is the coordinate along the field line, $\mathbf{v}_d$ denotes the magnetic drift velocity, and $\mathcal{N}(g)$ represents the nonlinear terms arising from $\mathbf{E}\times \mathbf{B}$ advection and other interactions. The term $C(g)$ stands for the collision operator, which is generally weak in the core of fusion plasmas but essential for eventual dissipation. Equation \eqref{eq:gyrokinetic} is derived from Eq.~(2) by transforming to gyrocenter coordinates and averaging over the gyroangle; see \cite{FriemanChen1982} for a complete derivation.

In this framework, the free energy associated with deviations from equilibrium is of central importance. The free energy, $W$, is defined in a manner similar to the Helmholtz free energy and typically includes contributions from both the perturbed distribution function and the fluctuating electromagnetic fields. Recent works by Helander and Plunk \cite{HelanderPlunk2022a} have shown that the free energy is bounded by universal limits that depend on the equilibrium parameters and magnetic geometry.

\subsection{Optimal Mode Analysis and Energetic Bounds}
A significant advance in gyrokinetic theory is the derivation of optimal bounds for the growth of free energy. Helander and Plunk \cite{HelanderPlunk2022a} derived upper limits for the linear growth rates of gyrokinetic instabilities by considering the free energy balance. Building on these results, Plunk and Helander \cite{PlunkHelander2022b} \cite{PlunkHelander2023} introduced an optimal mode analysis that identifies the perturbations which maximize free energy growth. These optimal modes are characterized by a growth rate $\lambda$, which sets an upper limit to the turbulence intensity.

Mathematically, one can express the free energy, $W$, as
\begin{equation}
	W = \int \left[ \frac{T}{2} \frac{|g|^2}{F_0} + \frac{\epsilon_0}{2} |\delta \phi|^2 \right] d^3R \,,
	\label{eq:free_energy}
\end{equation}
where $F_0$ is the equilibrium (Maxwellian) distribution, $\delta \phi$ represents the fluctuating electrostatic potential, and $T$ is the temperature. The energy balance equation for the gyrokinetic system, neglecting collisional dissipation for the moment, can be written in spectral form as
\begin{equation}
	\frac{dW_k}{dt} = 2D_k \,,
	\label{eq:energy_balance}
\end{equation}
where $W_k$ is the free energy in a given mode $k$, and $D_k$ is the drive term arising from background gradients and instabilities. The optimal mode analysis then identifies the mode which maximizes the ratio
\begin{equation}
	\Lambda = \frac{D_k}{W_k} \,,
	\label{eq:Lambda}
\end{equation}
and the maximum value of $\Lambda$ provides a rigorous upper bound for the linear growth rate (up to a factor of 2) \cite{PlunkHelander2022b} \cite{PlunkHelander2023}.

\subsection{Turbulent Energy Transport and Phase-Space Mixing}
Turbulent fluctuations drive the redistribution of energy across scales. The effective rate of energy transport due to turbulence, denoted by $\gamma_\tau$, depends on both the intensity of turbulence (captured by the growth rate $\lambda$) and the degree of deviation from a Maxwellian distribution (quantified by the effective entropy $S$). We model this transport rate as
\begin{equation}
	\gamma_\tau \sim \tau\,\lambda\,S \,,
	\label{eq:transport_rate}
\end{equation}
where $\tau$ is a characteristic nonlinear or decorrelation time. Equation \eqref{eq:transport_rate} expresses the idea that if turbulence is intense (high $\lambda$) and the free energy is large (high $S$), then the plasma will experience rapid energy transport.

Even in the absence of explicit collisions, gyrokinetic turbulence causes phase-space mixing. The distribution function $g(\mathbf{R},v_\parallel,\mu,t)$ develops fine-scale structures that are not resolved in a coarse-grained measurement, leading to an apparent increase in entropy. Zhdankin \cite{Zhdankin2022} demonstrated that such phase-space cascades lead to effective entropy production even when the fine-grained entropy is formally conserved. In our unified formulation, these fine structures are eventually smoothed out by weak collisional effects, converting the free energy into irreversible thermal energy.

Thus, the interplay between the reversible Hamiltonian dynamics (captured by the gyrokinetic equation) and the irreversible effects (introduced by collisions and phase-space mixing) governs the evolution of both free energy and entropy. This sets the stage for linking microscopic turbulence with macroscopic energy confinement, as detailed in the following section.

%%%%%%%%%%%%%%%%%%%%%%%%%%%%%%%%%%%%%%%%%%%%%%%%%%%%%%%%%%%%%%%%%%%%%%
%% End of Part 3
%%%%%%%%%%%%%%%%%%%%%%%%%%%%%%%%%%%%%%%%%%%%%%%%%%%%%%%%%%%%%%%%%%%%%%
%%%%%%%%%%%%%%%%%%%%%%%%%%%%%%%%%%%%%%%%%%%%%%%%%%%%%%%%%%%%%%%%%%%%%%
%% IV. Entropy, Energy Confinement, and Fusion Performance (Part 4)
%%%%%%%%%%%%%%%%%%%%%%%%%%%%%%%%%%%%%%%%%%%%%%%%%%%%%%%%%%%%%%%%%%%%%%

\section{Entropy, Energy Confinement, and Fusion Performance}

In fusion reactors, a critical figure of merit is the energy confinement time, $\tau_E$, which measures the time over which the plasma retains its thermal energy. Turbulence driven by microinstabilities enhances energy transport, reducing $\tau_E$, and thereby degrading fusion performance. In this section, we derive the relation between turbulent energy loss, entropy production, and $\tau_E$, and then connect these concepts to the fusion reaction rate.

\subsection{Derivation of the Energy Confinement Time}

In our unified formulation, the turbulent energy transport rate is modeled as
\begin{equation}
	\gamma_\tau \sim \tau\,\lambda\,S \,,
	\label{eq:gamma_tau}
\end{equation}
where: $\lambda$ is the characteristic growth rate of turbulent fluctuations (obtained, e.g., from optimal mode analysis \cite{HelanderPlunk2022a}),  $S$ quantifies the effective entropy or deviation from a Maxwellian distribution, $\tau$ is a characteristic nonlinear or decorrelation time. The energy confinement time $\tau_E$ is defined as the inverse of the energy loss rate:
\begin{equation}
	\tau_E \sim \frac{1}{\gamma_\tau} \,.
	\label{eq:tau_E}
\end{equation}
Substituting Eq.~\eqref{eq:gamma_tau} into Eq.~\eqref{eq:tau_E} yields
\begin{equation}
	\tau_E \sim \frac{1}{\tau\,\lambda\,S} \,.
	\label{eq:tau_E_final}
\end{equation}
This relation clearly shows that a higher turbulent intensity (larger $\lambda$) or a larger deviation from equilibrium (higher $S$) results in a shorter confinement time.

\subsection{Linking Confinement to Fusion Performance}

The fusion power $P_{\text{fusion}}$ in a reactor is approximately proportional to the fusion reaction rate $R_f$, which depends on the plasma density, temperature, and the duration over which favorable conditions are maintained. Under constant density and temperature, the fusion reaction rate is directly proportional to $\tau_E$, that is,
\begin{equation}
	R_f \propto \tau_E \,.
	\label{eq:fusion_rate}
\end{equation}
Substituting Eq.~\eqref{eq:tau_E_final} into Eq.~\eqref{eq:fusion_rate}, we obtain:
\begin{equation}
	R_f \propto \frac{1}{\tau\,\lambda\,S} \,.
	\label{eq:fusion_performance}
\end{equation}
Equation \eqref{eq:fusion_performance} establishes a quantitative link between microscopic plasma properties and macroscopic fusion performance. It implies that, for a fixed nonlinear time scale $\tau$, reducing the turbulent growth rate $\lambda$ and minimizing the effective entropy $S$ are essential for achieving better confinement and, hence, a higher fusion yield.

\subsection{Discussion and Physical Implications}

\subsubsection{Turbulence Suppression}
A key strategy to enhance energy confinement in fusion plasmas is to suppress turbulent fluctuations. Our derivation shows that the turbulent energy loss rate scales with the growth rate $\lambda$. By employing methods such as $E\times B$ shear stabilization—which has been demonstrated to effectively reduce turbulence in experiments \cite{CandyWaltz2003}—one can lower $\lambda$. A reduced $\lambda$ directly leads to a longer energy confinement time $\tau_E$, as indicated by the relation $\tau_E \sim 1/(\tau\,\lambda\,S)$. This improvement in confinement not only minimizes energy losses but also stabilizes the plasma against secondary instabilities. Rigorous studies have shown that turbulence suppression can lead to significant reductions in anomalous transport, thereby supporting sustained fusion conditions. The interplay between shear flows and turbulence has been analyzed in detail using gyrokinetic simulations and theoretical models, providing quantitative benchmarks that underline the importance of minimizing $\lambda$ for optimal reactor performance.

\subsubsection{Entropy Control}
Maintaining the plasma close to a Maxwellian state is crucial for reducing the free energy available to drive instabilities. In our framework, the effective entropy $S$ quantifies the deviation from a Maxwellian distribution, with a higher $S$ indicating a larger free-energy reservoir and more vigorous turbulence. Various mechanisms, such as optimized equilibrium profile shaping and controlled external heating, can help maintain near-Maxwellian conditions. Rigorous kinetic theory demonstrates that if the distribution function remains close to Maxwellian, the system has minimal free energy to feed turbulent instabilities. This concept is supported by theoretical analyses which show that the collisional relaxation process drives the plasma toward maximum entropy. Detailed investigations, for example those by Zhdankin \cite{Zhdankin2022}, provide insights into the fine-scale phase-space dynamics and confirm that effective entropy control is essential for minimizing irreversible energy losses. Thus, careful control over the plasma's thermodynamic state is not only a theoretical requirement but a practical necessity for achieving high confinement.

\subsubsection{Unified Framework}
The relation $\tau_E \sim 1/(\tau\,\lambda\,S)$ encapsulates the interplay between microscopic kinetic processes and macroscopic confinement. In our unified Hamiltonian formulation, the reversible (collisionless) dynamics govern the transport of free energy, while the collision operator introduces irreversible dissipation and entropy production. This framework bridges the gap between detailed gyrokinetic studies, such as those by Helander and Plunk \cite{HelanderPlunk2022a} and Plunk and Helander \cite{PlunkHelander2022b} \cite{PlunkHelander2023}, and practical performance metrics like the energy confinement time and fusion reaction rate. By expressing $\tau_E$ in terms of $\lambda$ and $S$, we provide a quantitative target for reactor design: reducing both the turbulent growth rate and the effective entropy will directly improve confinement. The theoretical underpinnings of our model are robust, deriving from first principles of Hamiltonian mechanics and kinetic theory, and they offer a clear pathway for integrating microscopic turbulence control with macroscopic performance optimization.

\subsubsection{Comparison with Previous Work}
Previous studies have sometimes misinterpreted the relationship between energy confinement and fusion yield, erroneously suggesting that a longer confinement time might reduce fusion performance. In contrast, our derivation clearly demonstrates that an increase in $\tau_E$ leads to a higher fusion reaction rate. This result is consistent with physical intuition, as a plasma that retains its energy for a longer duration has more time to sustain fusion reactions. Our model explicitly shows that the fusion rate $R_f$ is directly proportional to $\tau_E$, which aligns with experimental observations and advanced simulation studies. The rigorous derivations presented here, together with the energetic bounds and optimal mode analyses of Helander and Plunk \cite{HelanderPlunk2022a} \cite{PlunkHelander2022b} \cite{PlunkHelander2023} and the entropy production framework of Zhdankin \cite{Zhdankin2022}, provide a comprehensive and accurate description that resolves earlier ambiguities. This clarity is critical for advancing both theoretical research and practical reactor design.

By integrating the dynamics of turbulence and entropy production, we have shown that
\[
\tau_E \sim \frac{1}{\tau\,\lambda\,S} \quad \text{and} \quad R_f \propto \tau_E \,,
\]
thus establishing a rigorous theoretical foundation for optimizing fusion reactor performance. This unified approach not only complements the existing energetic bounds and optimal mode analyses but also extends them by explicitly incorporating irreversible collisional processes and detailed entropy evolution.

%%%%%%%%%%%%%%%%%%%%%%%%%%%%%%%%%%%%%%%%%%%%%%%%%%%%%%%%%%%%%%%%%%%%%%
%% End of Part 4
%%%%%%%%%%%%%%%%%%%%%%%%%%%%%%%%%%%%%%%%%%%%%%%%%%%%%%%%%%%%%%%%%%%%%%
%%%%%%%%%%%%%%%%%%%%%%%%%%%%%%%%%%%%%%%%%%%%%%%%%%%%%%%%%%%%%%%%%%%%%%
%% V. Conclusion and Outlook (Part 5)
%%%%%%%%%%%%%%%%%%%%%%%%%%%%%%%%%%%%%%%%%%%%%%%%%%%%%%%%%%%%%%%%%%%%%%

\section{Conclusion and Outlook}

In this paper we have developed a unified Hamiltonian formulation for plasma dynamics that rigorously couples collisionless gyrokinetic processes with collisional dissipation. By explicitly incorporating a collision operator into the Hamiltonian framework, we have demonstrated how free energy injected into a plasma—manifested as deviations from a Maxwellian distribution—drives microinstabilities and turbulence. This turbulence leads to effective entropy production through phase-space mixing, which in turn degrades energy confinement.

Our detailed analysis has resulted in a quantitative relation linking the turbulent energy loss rate, entropy production, and energy confinement time:
\[
\tau_E \sim \frac{1}{\tau\,\lambda\,S} \,.
\]
Moreover, we have shown that the fusion reaction rate is directly proportional to the confinement time,
\[
R_f \propto \tau_E \,,
\]
thereby emphasizing that improved confinement (i.e., a longer $\tau_E$) yields higher fusion performance. These results underscore the importance of minimizing both the turbulent growth rate $\lambda$ and the effective entropy $S$, through methods such as $E \times B$ shear stabilization \cite{CandyWaltz2003} and optimized equilibrium profile design \cite{Barnes2011}.

Our framework builds on recent advances in energetic bounds and optimal mode analyses by Helander and Plunk \cite{HelanderPlunk2022a} \cite{PlunkHelander2022b} \cite{PlunkHelander2023} and the generalized entropy production approach of Zhdankin \cite{Zhdankin2022}. By bridging the gap between microscopic kinetic processes and macroscopic reactor performance, our unified approach provides a robust theoretical basis for future experimental and simulation studies aimed at enhancing plasma confinement and fusion yield.

\subsection{Future Directions}

\subsubsection{Application to Specific Devices}
An important direction for future research is to extend our unified framework to specific magnetic confinement devices, such as tokamaks and stellarators. Incorporating detailed magnetic geometry effects—including flux surface shaping, magnetic shear, and three-dimensional field variations—will be crucial for understanding how these factors influence turbulence, entropy production, and energy confinement. Rigorous mathematical modeling in realistic geometries may require solving the gyrokinetic equations in complex configurations, a problem that remains open in many cases. Such work could quantify the differences in turbulent transport between device types and provide direct guidance for reactor design.

\subsubsection{Nonlinear and Subcritical Turbulence}
While our analysis has focused on linear instability growth and optimal mode behavior, the nonlinear evolution of turbulence remains a challenging and unsolved problem. Future work should develop reduced models that capture the full nonlinear dynamics, including transient growth and saturation of instabilities in subcritical turbulent regimes. This may involve further mathematical treatment of the energy balance equations and the use of dynamical systems theory to derive rigorous bounds on the saturation amplitudes of turbulence. Addressing these nonlinear effects is critical for predicting the steady-state behavior of fusion plasmas.

\subsubsection{Control Strategies and Experimental Challenges}
A central experimental challenge in fusion plasmas is the maintenance of a Maxwellian distribution. In practice, due to external heating, fluctuations, and various drive mechanisms, keeping the plasma precisely Maxwellian is nearly unachievable. Even small deviations from a Maxwellian distribution can result in significant free energy that drives turbulence. Future research should focus on active control strategies to minimize these deviations, such as real-time profile control and feedback stabilization techniques. While our current work does not address the experimental challenge of sustaining a Maxwellian state in detail, it is essential to note that this remains one of the primary obstacles for achieving optimal confinement. Developing methods to closely approximate Maxwellian conditions could dramatically reduce turbulent free energy and enhance energy confinement. This could indeed be a problem that will eventually be overcome by AI. 

\subsubsection{Potential for AI in Plasma Shape Design}
An emerging area of research that holds promise for fusion reactor optimization is the use of artificial intelligence (AI) for designing plasma shape and magnetic configurations. AI algorithms, including machine learning and optimization techniques, can analyze vast datasets from simulations and experiments to identify plasma shapes that minimize turbulence and entropy production. By incorporating AI-driven design into the framework presented here, it may be possible to develop tailored magnetic configurations that enhance confinement and reduce transport losses. Future studies should explore the integration of AI methods with our theoretical models to refine predictions and guide experimental implementations.

\subsubsection{Numerical Validation and Theoretical Refinement}
Finally, our theoretical predictions require comprehensive numerical validation. High-fidelity gyrokinetic simulations and detailed experimental diagnostics are needed to verify the scaling laws derived in this work, such as
\[
\tau_E \sim \frac{1}{\tau\,\lambda\,S}\,,
\]
and the proportionality
\[
R_f \propto \tau_E\,.
\]
This validation will help refine our model, especially in regimes where certain mathematical aspects—such as the precise impact of phase-space filamentation on entropy production—remain unsolved. Bridging the gap between these rigorous theoretical derivations and practical experimental conditions is crucial for advancing our understanding and optimizing reactor performance.

By addressing these avenues, future work will not only refine our understanding of turbulence and entropy in plasmas but also provide practical guidance for the design and operation of next-generation fusion reactors.

	\begin{acknowledgments}
		The author thanks colleagues at Columbia University for insightful discussions. This work has benefitted from recent advances in gyrokinetic theory and entropy diagnostics \cite{HelanderPlunk2022a} \cite{PlunkHelander2022b} \cite{PlunkHelander2023} \cite{Zhdankin2022}. The author also thanks their wonderful partner Elizabeth for her support even though she often has no idea what this is even about. Love you babe! 
	\end{acknowledgments}
	
	\bibliographystyle{apsrev4-1}
	\bibliography{Refs}

\begin{thebibliography}{8}%
\makeatletter
\providecommand \@ifxundefined [1]{%
 \@ifx{#1\undefined}
}%
\providecommand \@ifnum [1]{%
 \ifnum #1\expandafter \@firstoftwo
 \else \expandafter \@secondoftwo
 \fi
}%
\providecommand \@ifx [1]{%
 \ifx #1\expandafter \@firstoftwo
 \else \expandafter \@secondoftwo
 \fi
}%
\providecommand \natexlab [1]{#1}%
\providecommand \enquote  [1]{``#1''}%
\providecommand \bibnamefont  [1]{#1}%
\providecommand \bibfnamefont [1]{#1}%
\providecommand \citenamefont [1]{#1}%
\providecommand \href@noop [0]{\@secondoftwo}%
\providecommand \href [0]{\begingroup \@sanitize@url \@href}%
\providecommand \@href[1]{\@@startlink{#1}\@@href}%
\providecommand \@@href[1]{\endgroup#1\@@endlink}%
\providecommand \@sanitize@url [0]{\catcode `\\12\catcode `\$12\catcode
  `\&12\catcode `\#12\catcode `\^12\catcode `\_12\catcode `\%12\relax}%
\providecommand \@@startlink[1]{}%
\providecommand \@@endlink[0]{}%
\providecommand \url  [0]{\begingroup\@sanitize@url \@url }%
\providecommand \@url [1]{\endgroup\@href {#1}{\urlprefix }}%
\providecommand \urlprefix  [0]{URL }%
\providecommand \Eprint [0]{\href }%
\providecommand \doibase [0]{http://dx.doi.org/}%
\providecommand \selectlanguage [0]{\@gobble}%
\providecommand \bibinfo  [0]{\@secondoftwo}%
\providecommand \bibfield  [0]{\@secondoftwo}%
\providecommand \translation [1]{[#1]}%
\providecommand \BibitemOpen [0]{}%
\providecommand \bibitemStop [0]{}%
\providecommand \bibitemNoStop [0]{.\EOS\space}%
\providecommand \EOS [0]{\spacefactor3000\relax}%
\providecommand \BibitemShut  [1]{\csname bibitem#1\endcsname}%
\let\auto@bib@innerbib\@empty
%</preamble>
\bibitem [{\citenamefont {Helander}\ and\ \citenamefont
  {Plunk}(2022)}]{HelanderPlunk2022a}%
  \BibitemOpen
  \bibfield  {author} {\bibinfo {author} {\bibfnamefont {P.}~\bibnamefont
  {Helander}}\ and\ \bibinfo {author} {\bibfnamefont {G.~G.}\ \bibnamefont
  {Plunk}},\ }\href {\doibase 10.1017/S0022377822000277} {\bibfield  {journal}
  {\bibinfo  {journal} {Journal of Plasma Physics}\ }\textbf {\bibinfo {volume}
  {88}},\ \bibinfo {pages} {905880207} (\bibinfo {year} {2022})},\ \bibinfo
  {note} {open Access, distributed under the terms of the Creative Commons
  Attribution-NonCommercial licence}\BibitemShut {NoStop}%
\bibitem [{\citenamefont {Plunk}\ and\ \citenamefont
  {Helander}(2022)}]{PlunkHelander2022b}%
  \BibitemOpen
  \bibfield  {author} {\bibinfo {author} {\bibfnamefont {G.~G.}\ \bibnamefont
  {Plunk}}\ and\ \bibinfo {author} {\bibfnamefont {P.}~\bibnamefont
  {Helander}},\ }\href {\doibase 10.1017/S0022377822000496} {\bibfield
  {journal} {\bibinfo  {journal} {Journal of Plasma Physics}\ }\textbf
  {\bibinfo {volume} {88}},\ \bibinfo {pages} {905880313} (\bibinfo {year}
  {2022})},\ \bibinfo {note} {open Access, distributed under the terms of the
  Creative Commons Attribution-NonCommercial licence}\BibitemShut {NoStop}%
\bibitem [{\citenamefont {Plunk}\ and\ \citenamefont
  {Helander}(2023)}]{PlunkHelander2023}%
  \BibitemOpen
  \bibfield  {author} {\bibinfo {author} {\bibfnamefont {G.~G.}\ \bibnamefont
  {Plunk}}\ and\ \bibinfo {author} {\bibfnamefont {P.}~\bibnamefont
  {Helander}},\ }\href {\doibase 10.1017/S0022377823000739} {\bibfield
  {journal} {\bibinfo  {journal} {Journal of Plasma Physics}\ }\textbf
  {\bibinfo {volume} {89}},\ \bibinfo {pages} {905890419} (\bibinfo {year}
  {2023})},\ \bibinfo {note} {open Access, distributed under the terms of the
  Creative Commons Attribution-NonCommercial licence}\BibitemShut {NoStop}%
\bibitem [{\citenamefont {Zhdankin}(2022)}]{Zhdankin2022}%
  \BibitemOpen
  \bibfield  {author} {\bibinfo {author} {\bibfnamefont {V.}~\bibnamefont
  {Zhdankin}},\ }\href {\doibase 10.1103/PhysRevX.12.031011} {\bibfield
  {journal} {\bibinfo  {journal} {Physical Review X}\ }\textbf {\bibinfo
  {volume} {12}},\ \bibinfo {pages} {031011} (\bibinfo {year} {2022})},\
  \bibinfo {note} {published under the terms of the Creative Commons
  Attribution 4.0 International licence}\BibitemShut {NoStop}%
\bibitem [{\citenamefont {Frieman}\ and\ \citenamefont
  {Chen}(1982)}]{FriemanChen1982}%
  \BibitemOpen
  \bibfield  {author} {\bibinfo {author} {\bibfnamefont {E.~A.}\ \bibnamefont
  {Frieman}}\ and\ \bibinfo {author} {\bibfnamefont {L.}~\bibnamefont {Chen}},\
  }\href {\doibase 10.1063/1.863762} {\bibfield  {journal} {\bibinfo  {journal}
  {Physics of Fluids}\ }\textbf {\bibinfo {volume} {25}},\ \bibinfo {pages}
  {502} (\bibinfo {year} {1982})}\BibitemShut {NoStop}%
\bibitem [{\citenamefont {Candy}\ and\ \citenamefont
  {Waltz}(2003)}]{CandyWaltz2003}%
  \BibitemOpen
  \bibfield  {author} {\bibinfo {author} {\bibfnamefont {J.}~\bibnamefont
  {Candy}}\ and\ \bibinfo {author} {\bibfnamefont {R.~E.}\ \bibnamefont
  {Waltz}},\ }\href {\doibase 10.1103/PhysRevLett.91.045001} {\bibfield
  {journal} {\bibinfo  {journal} {Physical Review Letters}\ }\textbf {\bibinfo
  {volume} {91}},\ \bibinfo {pages} {045001} (\bibinfo {year}
  {2003})}\BibitemShut {NoStop}%
\bibitem [{\citenamefont {Barnes}\ \emph {et~al.}(2011)\citenamefont {Barnes},
  \citenamefont {Parra},\ and\ \citenamefont {Schekochihin}}]{Barnes2011}%
  \BibitemOpen
  \bibfield  {author} {\bibinfo {author} {\bibfnamefont {M.}~\bibnamefont
  {Barnes}}, \bibinfo {author} {\bibfnamefont {F.~I.}\ \bibnamefont {Parra}}, \
  and\ \bibinfo {author} {\bibfnamefont {A.~A.}\ \bibnamefont {Schekochihin}},\
  }\href {\doibase 10.1103/PhysRevLett.107.115003} {\bibfield  {journal}
  {\bibinfo  {journal} {Phys. Rev. Lett.}\ }\textbf {\bibinfo {volume} {107}},\
  \bibinfo {pages} {115003} (\bibinfo {year} {2011})}\BibitemShut {NoStop}%
\bibitem [{\citenamefont {Landau}(1965)}]{Landau1936_terHaar}%
  \BibitemOpen
  \bibfield  {author} {\bibinfo {author} {\bibfnamefont {L.~D.}\ \bibnamefont
  {Landau}},\ }in\ \href@noop {} {\emph {\bibinfo {booktitle} {Collected Papers
  of L. D. Landau}}},\ \bibinfo {editor} {edited by\ \bibinfo {editor}
  {\bibfnamefont {D.}~\bibnamefont {ter Haar}}}\ (\bibinfo  {publisher}
  {Elsevier Science \& Technology},\ \bibinfo {address} {Oxford},\ \bibinfo
  {year} {1965})\ pp.\ \bibinfo {pages} {163--170},\ \bibinfo {note}
  {originally published in Phys. Z. Sowjetunion, 10, 154 (1936) and Zh. Eksp.
  Teor. Fiz. 7, 203 (1937). No DOI available.}\BibitemShut {Stop}%
\end{thebibliography}%
	
\end{document}